\documentstyle[preprint,aps,12pt]{revtex}
\begin{document}
\draft
\preprint{YITP/K-1077}
\title{ Haldane's Fractional Exclusion Statistics \\
for Multicomponent Systems}
\author{ Takahiro Fukui and Norio Kawakami}
\address{Yukawa Institute for Theoretical Physics, Kyoto University,
Kyoto 606-01, Japan}
\date{4 August 1994}
\maketitle

\begin{abstract} % insert abstract here
The idea of fractional exclusion statistics proposed by
Haldane is applied to systems
with internal degrees of freedom, and its thermodynamics is examined.
In case of one dimension, various bulk quantities calculated
show that the critical behavior of such systems can be described by
$c=1$ conformal field theories
and conformal weights are completely characterized by statistical
interactions.
It is also found that statistical interactions
have intimate relationship with a topological order matrix
in Chern-Simons theory for the fractional quantum Hall effect.
\end{abstract}

%\narrowtext
% body of paper here

It is known that there are possibilities of existing
particles which obey fractional statistics in lower dimensions.
For example, in the fractional quantum Hall effect (FQHE)
quasi-particles carry fractional charges, which may be
described by Chern-Simons (CS) theory\cite{QHE}.
In this theory, as well as others,
the dimension of space plays a crucial role because
fractional statistics is defined on the basis of
monodromic properties of wavefunctions.
Recently, Haldane \cite{HAL} has proposed a new definition of
fractional statistics independent of the dimension of space.
It is based on the state-counting of
many-body systems and can be considered as a generalization of the
Pauli exclusion principle.  Imagine that there is a $N$-particle system
with species $\alpha_i, i=1,2,...,N$.
The $N$-body wavefunction can be expanded by
$i$th one-particle wavefunction by keeping
the coordinates of other particles fixed.
Let $D_{\alpha_i}$ be the dimension of such a basis.
By observing that $D_\alpha$ may be changed when a particle is added
without changing the boundary and the size of the system,
Haldane has introduced a statistical interaction $g_{\alpha\beta}$ defined by
\begin{equation}
\frac{\partial D_\alpha}{\partial N_\beta}=-g_{\alpha\beta},    \label{eq:SI}
\end{equation}
where $g_{\alpha\beta}$ is a constant independent of $\{N_\alpha\}$.
The special cases
of $g_{\alpha\beta}=0$ and $g_{\alpha\beta}=
\delta_{\alpha\beta}$ correspond to free
bosons and free fermions, respectively.
Since Haldane's work, several authors have investigated it in
detail \cite{MUR,WU,BER,NAY,BY}.

Wu \cite{WU} and Bernard and Wu \cite{BER}
have studied the thermodynamics of single-component
particles whose total number of states is given by
\begin{equation}
W=\prod_i\frac{(D_i+N_i-1)!}{N_i!(D_i-1)!} ,           \label{eq:NOS}
\end{equation}
with $D_i$ being a solution of eq.(\ref{eq:SI}),
where the species $\alpha$ is regarded as the index $i$ which labels the
momentum $k_i$.
They have shown that $g_{ij}$ is given by the two-body S matrix
for Bethe ansatz solvable models.
As simple examples in one dimension (1D),
they calculated $g_{ij}$ for the Bose gas with $\delta$-function interaction
and the Calogero-Sutherland model\cite{BER}. The former has non-zero
off-diagonal elements of $g_{ij}$ which they call mutual
statistics, while the latter has only diagonal ones,
$g_{ij}=g\delta_{ij}$. Nayak and Wilczek \cite{NAY} have called such
particles ``$g$-ons'' and examined low-temperature
properties in arbitrary dimensions.
It may be such $g$-ons that can be regarded as ideal particles obeying
fractional statistics.

In this Letter,
we will generalize this approach to systems with internal degrees of
freedom.
This has been briefly discussed in \cite{BER},
and a step has been made forward
in \cite{BY} for the 1D $U\rightarrow\infty$ Hubbard model.
First, we will obtain general formulae for thermodynamics,
and then calculate bulk quantities in case of 1D to find that
critical behavior of such systems can be
described by $c=1$ conformal field theories (CFT).
It will next be shown that there is intimate relationship
between statistical interactions and a topological order
matrix in CS theory for the FQHE\cite{WEN}.

Let us consider a system of $N$ particles whose states are
classified by the labels $\alpha$ and $i$, where
$\alpha ~(=1,2,...,M)$ and $i$ correspond to
internal degrees of freedom and the momentum of particles, respectively.
Denoting explicitly these two kinds of indices
in eq.(\ref{eq:SI}), the statistical interaction $g_{\alpha i\beta j}$
for the multicomponent system reads,
\begin{equation}
\frac{\partial D_{\alpha i}}{\partial N_{\beta j}}=
-g_{\alpha i\beta j},                                     \label{eq:GSI}
\end{equation}
which results in
\begin{equation}
D_{\alpha i}=-\sum_{\beta j}g_{\alpha i\beta j}N_{\beta j}+D_{\alpha
i}^0 ,                                                     \label{eq:DI}
\end{equation}
where $N_{\alpha i}$ ($D_{\alpha i}$) is the number of particles
(holes) labeled by $\alpha i$,
and $D_{\alpha i}^0$ is the number of holes when
there are no particles in the systems.
Suppose that we have no particles
in the system under consideration. Then there are only
$M$ kinds of free holes. Therefore, it is natural to choose
$D_{\alpha i}^0=D_\alpha^0$ with $D_\alpha^0=D^0$, which we will call a
symmetric basis in this paper.
Another important choice is $D_\alpha^0=D^0\delta_{\alpha 1}$, namely,
$\alpha =1$ and $\alpha >1$ classify charge and internal degrees of freedom,
respectively. This is referred to as a  hierarchical basis.
It is noteworthy that one can find the same bases also
in a CS theory for the FQHE\cite{WEN},
as we shall see later.
The distribution functions of particles and holes
labeled by $\alpha i$ are defined as
\begin{equation}
\rho_{\alpha i}\equiv\frac{N_{\alpha i}}{D^0}, \quad
\rho_{\alpha i}^{(h)}\equiv\frac{D_{\alpha i}}{D^0} .\label{eq:DEFR}
\end{equation}
Charge for each component is also defined by
$t_\alpha\equiv D_\alpha^0/D^0$,
which reduces to $\vec t=(1,1,..,1)$ and $(1,0,...,0)$ in
the symmetric and the hierarchical bases, respectively.
Note that the quantities (\ref{eq:DEFR}) play a key role in the
fractional exclusion statistics because they are generalizations of
Fermi (or Bose) distribution function.
In the following, we consider multicomponent systems whose
total energy is given by the summation of bare single-particle
energies, $E=\sum_{\alpha i}\epsilon_{\alpha i}^0N_{\alpha i}$,
where we set $\epsilon_{\alpha i}^0\equiv\epsilon_i^0t_\alpha$.
It should be noted that many-body effects are in general
incorporated in the distribution of $N_{\alpha i}$.

Let us now examine the thermodynamics of a system whose number of states
is given by eq.(\ref{eq:NOS}) with replacing $i\rightarrow\alpha i$.
Notice that eq.(\ref{eq:DI}) is rewritten as
\begin{equation}
\rho_{\alpha i}^{(h)}+\sum_{\beta j}g_{\alpha i\beta j}\rho_{\beta j}=
t_\alpha .                                           \label{eq:BAE}
\end{equation}
It is now straightforward to investigate the thermodynamics following
the method of Wu\cite{WU} and Bernard and Wu\cite{BER}, which is based on
Yang and Yang's idea \cite{YANG}.
Thermal equilibrium is realized by minimizing the
thermodynamic potential
$\Omega\equiv E-TS-\sum_\alpha\mu_\alpha N_\alpha$,
where $N_\alpha\equiv\sum_iN_{\alpha i}$ and $\mu_\alpha$ denote
the total number of particles and the chemical potential
for the species $\alpha$, respectively.
The entropy $S$ is given by $S=\ln W$ with $W$ replaced
$i\rightarrow \alpha i$ in eq.(\ref{eq:NOS}).
The results are as follows. The free energy $F$
at equilibrium is
\begin{equation}
F=\sum_\alpha\mu_\alpha N_\alpha
-T\sum_{\alpha i}D_\alpha^0\ln
[1+\exp (-\epsilon_{\alpha i}/T)] ,                            \label{eq:FE}
\end{equation}
where the ``dressed energy'' $\epsilon_{\alpha i}=T\ln (D_{\alpha i}/
N_{\alpha i})=T\ln(\rho_{\alpha i}^{(h)}/\rho_{\alpha i})$ is determined by
\begin{equation}
\epsilon_{\alpha i}-T\sum_{\beta j}\widetilde g_{\beta j\alpha i}\ln [1+\exp
(-\epsilon_{\beta j}/T)] =\epsilon_{\alpha i}^0-\mu_\alpha ,    \label{eq:DE}
\end{equation}
where $\widetilde g_{\alpha i\beta j}\equiv g_{\alpha i\beta j}-
\delta_{\alpha i\beta j}$.
Though we here assume that there is no degeneracy for each state,
we can also formulate the thermodynamics by introducing the
corresponding weight factor in eqs.(\ref{eq:FE}) and (\ref{eq:DE})
when each state has a
certain degeneracy.
The formulae (\ref{eq:FE}) and (\ref{eq:DE}) describe thermodynamics of
multicomponent systems obeying the fractional exclusion statistics
(\ref{eq:SI}).

Now consider low-temperature properties
in the thermodynamic limit, which is defined by
$N_\alpha\rightarrow\infty$ and $D^0\rightarrow\infty$ proportionally
\cite{BER}. Hereafter, we restrict ourselves to 1D systems for simplicity.
Then eqs.(\ref{eq:BAE})$\sim$(\ref{eq:DE}) read
\begin{eqnarray}
&&\rho_\alpha^{(t)}(k_\alpha )=t_\alpha -\sum_\beta
\int_{-\infty}^\infty
dk_\beta\widetilde g_{\alpha\beta}(k_\alpha ,k_\beta
)\rho_\beta (k_\beta ),     \label{eq:BAET} \\
&&F=\sum_\alpha\mu_\alpha N_\alpha-T\sum_\alpha D_\alpha^0
\int_{-\infty}^\infty dk_\alpha\ln
[1+\exp (-\epsilon_\alpha (k_\alpha )/T)] ,             \label{eq:FET}\\
&&\epsilon_\alpha (k_\alpha )=\epsilon_\alpha^0(k_\alpha )-\mu_\alpha
+T\sum_\beta\int_{-\infty}^\infty dk_\beta\widetilde g_{\beta\alpha}
(k_\beta ,k_\alpha )\ln [1+\exp
(-\epsilon_\beta (k_\beta )/T)]                               \label{eq:DET},
\end{eqnarray}
where $\rho_\alpha^{(t)}\equiv\rho_\alpha +\rho_\alpha^{(h)}$.
The total energy is given by $e\equiv E/D^0=\sum_\alpha e_\alpha$ with
$e_\alpha =\int_{-\infty}^\infty dk_\alpha\epsilon_\alpha (k_\alpha
)\rho_\alpha (k_\alpha)$.
Assume $\widetilde g_{\alpha\beta}(-k_\alpha ,-k_\beta)=
\widetilde g_{\alpha\beta}(k_\alpha ,k_\beta )$
and $\epsilon_\alpha^0(-k_\alpha )=\epsilon_\alpha^0(k_\alpha )$.
Then at $T=0$ there exists a ``generalized Fermi level''
$Q_\alpha$ which satisfies $\epsilon_\alpha (k_\alpha )<0$
for $|k_\alpha |<Q_\alpha$
and $\epsilon_\alpha (k_\alpha )>0$ for $|k_\alpha |>Q_\alpha$.
Hence $\rho_\alpha (k_\alpha )=0$ for $|k_\alpha |>Q_\alpha$,
and $\rho_\alpha (k_\alpha )$ for $|k_\alpha |<Q_\alpha$
and $\epsilon_\alpha (k_\alpha )$ are given by the solutions of
the following equations:
\begin{eqnarray}
&&\rho_\alpha (k_\alpha)=t_\alpha -\sum_\beta
\int_{-Q_\beta}^{Q_\beta}dk_\beta\widetilde g_{\alpha\beta}
(k_\alpha ,k_\beta )\rho_\beta (k_\beta ) ,\\
&&\epsilon_\alpha (k_\alpha )=\epsilon_\alpha^0(k_\alpha )
-\mu_\alpha -\sum_\beta
\int_{-Q_\beta}^{Q_\beta}dk_\beta\epsilon_\beta (k_\beta )\widetilde
g_{\beta\alpha}(k_\beta ,k_\alpha ) .\label{eq:DEZ}
\end{eqnarray}
Therefore, in the low-temperature limit, eq.(\ref{eq:FET}) is
expanded as
\begin{equation}
F(T)=F(T=0)-\frac{\pi}{6}T^2(2\pi D^0)\sum_\alpha \frac{1}{v_\alpha}
\end{equation}
in terms of the Fermi velocity $v_\alpha\equiv\epsilon_\alpha '(Q_\alpha )/
\rho_\alpha (Q_\alpha )$.
The specific heat in low-temperatures in unit length
is then given by $C/(TL)=(\pi /3)\sum_\alpha (1/v_\alpha )$, where
$D^0=L/(2\pi)$ is used\cite{BER}.
Hence, the critical behavior of 1D systems whose statistical interaction is
defined by eq.(\ref{eq:GSI}) can
be described by $M$ independent $c=1$ CFT\cite{BLO}.
The responses to ``external fields'' $\vec\mu$ are also calculated as
\begin{equation}
\chi_{\alpha\beta}\equiv\frac{\partial n_\alpha}{\partial \mu_\beta}
=2({\cal Z}V^{-1}{\cal Z}^T)_{\alpha\beta} ,       \label{eq:CHI}
\end{equation}
where $n_\alpha\equiv N_\alpha /D^0=\int_{-\infty}^\infty
dk_\alpha\rho_\alpha (k_\alpha )$,
${\cal Z}_{\alpha\beta}\equiv
Z_{\alpha\beta}(Q_\beta )$ is a boundary-valued ``dressed charge''
$Z_{\alpha\beta}(k_\beta )$ defined by\cite{IZ}
\begin{equation}
Z_{\alpha\beta}(k_\beta )=\delta_{\alpha\beta}-\sum_\gamma
\int_{-Q_\gamma}^{Q_\gamma}dk_\gamma Z_{\alpha\gamma}(k_\gamma )
\widetilde g_{\gamma\beta}(k_\gamma ,k_\beta) ,
\end{equation}
and a matrix $V$ is defined by
$V_{\alpha\beta}\equiv v_\alpha\delta_{\alpha\beta}$.
{}From eq.(\ref{eq:CHI}), the compressibility can be obtained as
$\chi_c=\sum_{\alpha\beta}t_\alpha\chi_{\alpha\beta}t_\beta$.
The excitation spectrum in CFT limit is classified as\cite{IZ}
\begin{equation}
\Delta e=\frac{1}{4}\vec{\Delta n} ({\cal Z}^{-1})^TV{\cal Z}^{-1}
\vec{\Delta n}+\vec{\Delta d}{\cal Z}V{\cal Z}^T\vec{\Delta d}
\label{eq:FSS}
\end{equation}
except for particle-hole excitations,
where $\vec{\Delta n}$ ($\vec{\Delta d}$) is a $M$-component
vector of quantum numbers which change the number of particles (carry
the large momentum).
This expression also confirms that the critical behavior is described
by $M$ independent $c=1$ gaussian CFT.

Let us now consider a simple system with
\begin{equation}
 g_{\alpha i\beta j}=
\delta_{ij}G_{\alpha\beta}.    \label{eq:SCG}
\end{equation}
The choice (\ref{eq:SCG}) implies that
the interaction in the momentum space is local, i.e., it takes a
$\delta$-function form.
We refer to particles obeying (\ref{eq:SCG}) as
{\it ideal multicomponent gases with fractional exclusion statistics which is
characterized by the statistical interaction $G_{\alpha\beta}$}.
One can indeed find that eqs.(\ref{eq:BAE}) and (\ref{eq:BAET}) reduce to
$\rho_\alpha^{(h)}(k)+\sum_\beta G_{\alpha\beta}\rho_\beta (k)=t_\alpha$,
which shows an asymmetry between particles and holes as discussed
in refs.\cite{SUTHL,BER,HA}.
We will later discuss concrete models in 1D which show
fractional exclusion statistics (\ref{eq:SCG}).
{}From eq.(\ref{eq:SCG}), the distribution
of particles with species $\alpha$ at finite temperature is given by
\begin{equation}
\rho_\alpha (k)=\sum_\beta
G_{\alpha\beta}^{-1}(k,T)t_\beta ,\label{eq:FTR}
\end{equation}
where $G_{\alpha\beta}(k,T)\equiv
\exp (\epsilon_\alpha (k)/T)\delta_{\alpha\beta}+G_{\alpha\beta}$,
and $\epsilon_\alpha (k)$ is a solution of eq.(\ref{eq:DET}), i.e.,
$\prod_\beta (e^{\epsilon_\beta /T})^{G_{\beta\alpha}}
(1+e^{\epsilon_\beta /T})^{\delta_{\beta\alpha}-G_{\beta\alpha}}
=e^{(\epsilon_\beta^0 -\mu_\beta )/T}$.
We emphasize that the formula (\ref{eq:FTR}) is a generalization of the
Fermi (or Bose) distribution function to a multicomponent ideal gas
with the statistical interaction $G_{\alpha\beta}$.
We here show simple examples in the symmetric basis.
First, when $G_{\alpha\beta}=g$, distribution in
eq.(\ref{eq:FTR}) becomes that of ideal fractional one-component particles,
$\rho =\rho^{(g)}$, which has been discussed in \cite{HAL,MUR,BER,NAY}.
This formula further reduces to $1/(e^{(\epsilon^0-\mu )/T}-1)$ for
free bosons ($g=0$) and to $1/(e^{(\epsilon^0-\mu )/T}+1)$ for
free fermions ($g=1$).
As another simple example of two-component system with
$G_{11}=G_{22}=g$ and $G_{12}=G_{21}=f$, we have
$\vec\rho =(\rho^{(g')},\rho^{(g')})$ with $g'\equiv g+f$.

Consider now the $T=0$ case with general $G_{\alpha\beta}$.
There appear $M$ kinds of ``generalized Fermi levels''
$Q_1,Q_2,...,Q_M$. Define the ordering of
$\alpha$ by $\tau (\alpha )$, such that
$Q_{\tau (1)}\ge Q_{\tau (2)}\ge ...\ge Q_{\tau (M)}$.
Then, in the region
$Q_{\tau (n)}> |k|> Q_{\tau (n+1)}$, we find
$\rho_{\tau (\alpha )}(k)=\sum_{\beta =1}^n\tau_n
(G)^{-1}_{\alpha\beta}t_{\tau (\beta )}$ for $\alpha =1,...,n$ and
$\rho_{\tau (\alpha )}(k)=0$ for $\alpha =n+1,...,M$,
where $\tau_n (G)$ is a
$n\times n$ matrix whose elements are given by $\tau_n
(G)_{\alpha\beta}=G_{\tau (\alpha )\tau (\beta )}$ with $\alpha ,\beta
=1,2,...,n$.
In the absence of external fields
one finds, by putting $\mu_\alpha =\mu t_\alpha$,
that the ground state is obtained by setting
$Q_1=Q_2=...=Q_M\equiv Q$. The dressed energy $\epsilon_\alpha (k)$,
distribution function $\rho_\alpha$
and the density of particles
$n_c\equiv N_c/D^0$, where $N_c\equiv\sum_\alpha
N_\alpha t_\alpha$, are calculated as
\begin{eqnarray}
&&\epsilon_\alpha (k)=\sum_\beta G^{-1}_{\alpha\beta}
(\epsilon_\beta^0(k)-\mu_\beta ) ,                    \label{eq:DES}\\
&&\rho_\alpha=\sum_\beta G^{-1}_{\alpha\beta}t_\beta ,\label{eq:RHO}\\
&&n_c=2Q\nu, \quad \nu\equiv \sum_{\alpha\beta}t_\alpha
G^{-1}_{\alpha\beta}t_\beta  .                            \label{eq:FF}
\end{eqnarray}
%%%%%%%%%%%%%%%%%%%%%%%% Kawakami %%%%%%%%%%%%%%%%%%%%%%%%
These equations show that Fermi velocities are
calculated as $v_\alpha =Q=n_c/(2\nu)$,
which are independent of their components.
If we take $\epsilon^0(k)=k^2/2$ as the bare spectrum,
we then find the ground state energy $e_g=n_c^3/(24\nu^2)$ and
the compressibility $\chi_c=4\nu^2/n_c$.
Hence, these bulk quantities are determined solely by
$\nu$ which does not involve full information of the matrix $G$.
However, if we observe the excitation spectrum
$\Delta e=(v/4)\vec{\Delta n}G\vec{\Delta n}+v\vec{\Delta
d}G^{-1}\vec{\Delta d}$,
there shows up every element of the matrix $G$, owing to the fact that
$M$ kinds of Fermi velocities are the same (See eq.(\ref{eq:FSS})).
Here we assume $G^T=G$.
Note that this formula for the excitation is of a desirable form
required by CFT, and we thus find conformal dimensions
determined completely by the
matrix $G$ following  the idea of finite size-scaling\cite{BLO}.

A remarkable point is that the above matrix $G$
of statistical interactions
has close relationship to a topological order matrix in the FQHE.
To see this explicitly, it is
more convenient to take $\epsilon^0(k)=k$ as
the bare spectrum, which only contains right-going
particles. It is known that such {\it chiral}
particles in 1D may describe essential properties
for edge states of the FQHE \cite{EDGE}.
In this case, the excitation spectrum of the
present system has the form $\Delta e=(v/4)\vec{\Delta n}G\vec{\Delta n}$,
which may be described by the holomorphic piece of
CFT. Recall here that a certain hierarchy of the FQHE
can be described by a multicomponent
CS Lagrangian ${\cal L}=
\epsilon^{\mu\nu\lambda} a_{\alpha\mu}K_{\alpha\beta}
\partial_\nu a_{\beta\lambda}$ in which
the construction of a hierarchy is characterized by
the {\it topological order matrix} $K$ \cite{WEN}.
An important point is that the excitation spectrum
obtained here for the chiral model coincides with
that for edge states\cite{EDGE} characterized by
the topological order matrix $ K $\cite{EXC}.
Therefore we arrive at an instructive
consequence that  the matrix $G$ of statistical
interaction can be identified with
a topological order matrix $ K$ in CS theory for
the FQHE.

For example, the topological-order matrix $K$ used in ref.\cite{WEN,CHU}
is written in the symmetric basis,
%%%%%%%%%%%%%%%%%%%%
\begin{equation}
K=\sum_{\alpha =1}^M(p_\alpha -2+\delta_{\alpha 1})
C_M^{(\alpha )}+I  \label{eq:SK}
\end{equation}
%%%%%%%%%%%%%%%%%%%%
with an odd integer $p_1$ and even integers $p_\alpha$
($\alpha >1$), where $I$ is a unit matrix
and $C_M^{(\alpha )}$ is a $M\times M$ matrix
defined by $(C_M^{(\alpha )})_{ij}=1$
for $i,j=\alpha ,\alpha +1,...,M$ and
$=0$ for otherwise.
Substituting eq.(\ref{eq:SK}) into
eq.(\ref{eq:FF}), we have
%%%%%%%%%%%%%%%%%%%%%%%%%%%%%%%%%%%%%%%%%%%
\begin{equation}
\nu =\frac{1}{p_1-\displaystyle{\frac{1}{p_2-\cdots
\displaystyle{\frac{1}{p_M}}}}}         .        \label{eq:FFE}
\end{equation}
%%%%%%%%%%%%%%%%%%%%%%%%%
Note that this quantity is nothing but the
filling factor for the FQHE described by (\ref{eq:SK}).
It is known\cite{WEN} that
the symmetric basis is converted into the hierarchical basis by the
matrix $W$ defined by
$W_{\alpha\beta}=\delta_{\alpha\beta}-\delta_{\alpha+1,\beta}$ as
follows; $\vec t\rightarrow\vec tW$ and $K\rightarrow W^TKW$.
In this basis, distribution functions (\ref{eq:RHO}) are
easily calculated at zero temperature;
$\rho_\alpha =\prod_{\beta =1}^\alpha\nu_\beta$, where
$\nu_\beta =\sum_{i=1}^\beta G^{(\beta )-1}_{i1}$ with
$\beta\times\beta$ matrix $G^{(\beta )}$
defined by $G_{ij}^{(\beta)}\equiv G_{M-\beta +i,M-\beta +j}$.
In particular,
when $p_1=2m+1$ and $p_\alpha =2$ for $\alpha =2,3,...,M$,
we have the filling factor $\nu =M/(2mM+1)$ and
the corresponding distribution functions $\vec\rho =(\nu,
(M-1)/M,(M-2)/(M-1),...,2/3,1/2)$.  This family corresponds to
a fundamental hierarchy which has been observed in the FQHE\cite{QHE}.

Finally  some comments are in order for  concrete models in 1D,
which exhibit essential properties of fractional exclusion statistics.
It has been claimed that {\it ideal} particles with
fractional exclusion statistics are realized  by 1D
quantum models with $1/r^2$ interaction\cite{ISMD,ISMS} and
their multicomponent versions\cite{ISMM}.
In fact, this fractional property has been shown explicitly for the
Haldane-Shastry spin chain \cite{HSF} and also for the
Calogero-Sutherland model\cite{SUTHL,HA,CSF}.
For cases with internal degrees of freedom
discussed here, we find that eqs.(\ref{eq:BAE})$\sim$(\ref{eq:DE})
indeed reduce to the Bethe ansatz equations
for multicomponent Calogero-Sutherland
model with a special coupling\cite{KAWA},
if we convert the expressions to those of the hierarchical basis.
Hence, an ideal multicomponent gas with statistical interaction
$G$ may also be realized by multicomponent $1/r^2$ models.

On the other hand, for interacting 1D particles with general type of
interaction, excitations cannot be regarded as
ideal particles. Nevertheless,
if we restrict ourselves to the CFT (low-energy)
limit of such systems, the excitation  can be well approximated
by ideal particles with fractional exclusion statistics.
 For example,  consider the 1D Hubbard model
in the CFT limit. There are two elementary modes called spinon and holon,
which have different velocities $v_s\ne v_h$ in general. Hence,
it seems difficult to deduce the matrix $G$
from eq.(\ref{eq:FSS}). However, these velocities
are not universal, and therefore by choosing $v_s=v_h$
hypothetically, we can conclude that statistical
interaction $G$ is given by
$G=I+\alpha C_2^{(1)}$ with $\alpha\equiv (1/K_\rho-1)/2$, where
$K_\rho$ ($1/2 \leq K_\rho \leq 1$) is the
critical exponent (divided by 4) for the
$4k_F$ oscillating part of the density-density
correlation function\cite{HUB}.  This kind of interpretation for the
elementary excitation is valid in general for the class of
multicomponent Luttinger liquids.

In summary, we have applied the idea of Haldane's fractional statistics
to systems with internal degrees of freedom.
The critical behavior of such 1D systems can be
described by multiple $c=1$ CFT with conformal weights
given by statistical interaction $G$.
It has also been shown that there is intimate relationship
between statistical interactions and a topological order
matrix in CS theory for the FQHE.
It is quite interesting and still open to
study how the fractional exclusion statistics
control the physics of correlated particles in higher dimensions.

This work is supported by the Grant-in-Aid from the Ministry of
Education, Science and Culture.


\begin{references}
\bibitem{EXC} Except for velocities which are non-universal quantities
depending on a shape of FQHE droplet.
\end{references}
\end{document}